\documentclass[11pt]{article}
\pdfoutput=1
\usepackage[english]{babel}
\usepackage[utf8]{inputenc}
\usepackage{xcolor} 

\usepackage[T1]{fontenc}
\usepackage{fullpage}
\usepackage{cite}

\usepackage[activate={true,nocompatibility},final,tracking=true,kerning=true,spacing=true]{microtype}
\microtypecontext{spacing=nonfrench}

\usepackage{amssymb}
\usepackage{amsmath}

\usepackage{hyperref}

\usepackage[ruled]{algorithm2e}
\usepackage[noend]{algorithmic}
\usepackage{array}
\usepackage{multirow}
\usepackage{nicefrac}

\usepackage{pgfplots, pgfplotstable}
\usepackage{tikz}
\usetikzlibrary{patterns}
\usetikzlibrary{fit}
\usetikzlibrary{external}
\usepgfplotslibrary{dateplot}
\usepackage[width=.75\textwidth]{caption}

\makeatletter
\newcommand{\Id}[1]{{\sf #1}}
\newcommand{\unit}[1]{\ensuremath{\, \mathrm{#1}}}
\newcommand{\unitfrac}[2]{\ensuremath{\, \mathrm{\nicefrac{#1}{#2}}}}
\newcommand\footnoteref[1]{\protected@xdef\@thefnmark{\ref{#1}}\@footnotemark}

\newcounter{reviewcnt}
\newcommand{\review}[1]{{\scriptsize{\bf{\,(\arabic{reviewcnt})\,#1\addtocounter{reviewcnt}{1}}}}}

\newcommand{\frage}[1]{{\color{blue}\scriptsize{\bf{#1}}}}
\newcommand{\antwort}[1]{{\color{red}\,\scriptsize{\emph{#1}}}}
\renewcommand{\frage}[1]{}
\renewcommand{\antwort}[1]{}
\renewcommand{\review}[1]{}
\newcommand{\mycomment}[1]{}
\makeatother

\begin{document}
%
\title{Fast OLAP Query Execution in Main Memory on Large Data in a Cluster}
\author{Demian Hespe$^*$ \\ \href{mailto:hespe@kit.edu}{hespe@kit.edu} \and
  Martin Weidner$^{\dagger}$ \\ \href{mailto:weidner@sap.com}{weidner@sap.com}
  \and Jonathan Dees$^{\dagger}$ \\ \href{mailto:dees@sap.com}{dees@sap.com}
  \and Peter Sanders$^*$ \\ \href{mailto:sanders@kit.edu}{sanders@kit.edu} \and \\ $^*$ Karlsruhe Institute of Technology, Karlsruhe, Germany\\ $^\dagger$ SAP SE, Walldorf, Germany}
\maketitle


\begin{abstract}
Main memory column-stores have proven to be efficient for processing analytical queries.
Still, there has been much less work in the context of clusters.
Using only a single machine poses several restrictions:
Processing power and data volume are bounded to the number of cores and main memory fitting on one tightly coupled system. 
To enable the processing of larger data sets, switching to a cluster becomes necessary.
In this work, we explore techniques for efficient execution of analytical SQL queries on large amounts of data in a parallel database cluster while making maximal use of the available hardware.
This includes precompiled query plans for efficient CPU utilization, full parallelization on single nodes and across the cluster, and efficient inter-node communication.
We implement all features in a prototype for running a subset of TPC-H benchmark queries.
We evaluate our implementation using a 128 node cluster running TPC-H queries with 30\,000 gigabyte of uncompressed data.

\end{abstract}

\section{Introduction}
Today, main memory column-stores are widely used for the efficient execution of analytical queries and lead to significant performance advantages \cite{abadi2008column}.
While the performance of the database systems increases, challenges increase too as users always want lower query response times and process bigger data.
There are two approaches to improve performance and data size \cite{valduriez1993parallel}:

\begin{itemize}
\item scale-up: improve the power of a single machine
\item scale-out: use multiple connected machines
\end{itemize}



 A scale-up solution improves the capabilities of a single machine.
 As current speed improvements of single processing units are getting smaller, today, extra processing power is usually gained by adding more processing cores into a tightly integrated processor.
 To use the full processing power on a single machine, parallelization of query processing functions becomes obligatory.
 Here, it is important to cover all major processing steps since each serially executed part can quickly become the bottleneck of execution and prevent satisfying speedup factors \cite{gustafson1988reevaluating}.
 Although the number of cores on a single chip is constantly increasing, the current architecture usually contains an upper limit.
 The same holds for main memory, even if current main memories with several terabytes per machine appear already to be quite large.
 To go beyond these limits and process several \emph{peta}bytes of data with sufficient processing power requires more than one machine.

A scale-out solution includes the use of several machines working together, usually across an efficient network.
 Using several machines can not only exceed the maximum processable data volumes of a single machine but can also be very cost competitive.
 Several small machines providing performance equal to one high-end server might often be significantly cheaper, making it an attractive alternative.

 A cluster can be a very efficient solution, but it can also increase system complexity significantly.
 In particular, if each computation node in the cluster holds only parts of the overall data, an efficient communication network between the nodes is required.
 In a shared-disk implementation, all nodes in a cluster can access the same logical disk such that every node can access all the data on disk.
 However, the shared disk can become a new bottleneck and reduce the main memory advantages.
 Indeed, we focus on shared-nothing systems within a cluster, i.e., we have no shared resource and every node stores only parts of the overall data, making large data volumes possible.
 On a single shared-memory node it is essential to utilize all available threads and perform parallel execution.
 Inter-query parallelism is not enough if the number of concurrent queries is lower than the number of threads.
 It is also not enough when we just want to minimize the execution time of a single query running alone on the system.
 Therefore, parallel execution within query operators (intra-operator) or between operators (inter-operator) is mandatory in order to use the full hardware capacity.

Besides parallel databases, no-SQL alternatives like MapReduce have attracted increasing attention for processing large amounts of data in recent years. However, it should be noted that for complex queries, SQL is both higher level and allows faster processing in many situations \cite{pavlo09}.

In this work, we explore with which techniques and algorithms we can achieve maximal performance for executing analytical SQL queries in a distributed database cluster.
In particular, we combine major techniques relevant for performance in one
system to make maximal use of the available hardware. In order to improve
performance of queries that only return a human-readable part of the complete
result, we also develop new algorithms for top-k selection.
Aiming for a shared-nothing system, we want to support large amounts of data that fit into the overall main memory of the cluster system.
Using a subset of different analytical queries from the popular TPC-H benchmark, we study the most efficient way for execution with the available hardware, trying to reach a new performance baseline.
We contribute a performance study that includes a distributed implementation evaluated in a networked cluster with up to 128 nodes.
While analyzing optimization possibilities for these TPC-H queries in detail, we are looking for solutions to systematically apply our optimizations and communication patterns on arbitrary queries.
At the end, we compare our implementation with the current official \mbox{TPC-H} record holders.

In order to make full use of the available hardware in our cluster, we combine the following principles:

\begin{itemize}
\item efficient single core data processing
\item full parallelization on a single machine
\item efficient distributed execution and communication 
\end{itemize}

Note that distributed query execution usually increases the number of query passes.
Additional communication steps during query execution are required to request data or to ship data to a remote node for further processing because in a shared-nothing environment, related data resides often on a different node.
Most techniques for distributed query execution are orthogonal to the local query execution model.
Hence, these techniques can also be applied in different execution models.

Exchanging data between nodes in a cluster requires efficient
inter-node communication.  We use the message passing
library MPI \cite{gropp1999using} for this purpose.  MPI provides
advanced collective operations like gather, (personalized) all-to-all,
or reduction. Theses collective operations enable efficient and
scalable communication between all nodes in the network. Relying only
on point-to-point communication could easily introduce communication
bottlenecks as scalable communication algorithms are often
non-trivial. The possibility of adding custom data types and reduction
functions to the MPI operators further improves performance.

\mycomment{As a further contribution, we developed a classification for SQL queries that are executed within a distributed environment.
In terms of data distribution and partitioning, the classes take into consideration the locality of processed tuples.
The classification can be applied to all possible SQL queries and defines a strategy for execution.}

The structure of the paper is as follows.
First, we discuss related work.
\mycomment{was: Second, we extend and refine an existing classification for the distributed query execution.
In this context, we also engineer efficient solutions to reduce communication costs.}
Second, we engineer efficient solutions to reduce communication costs.
Third, we discuss our methods in the context of the TPC-H benchmark. 
Next, we evaluate our implementation experimentally.
Finally, we summarize the key results of our work and sketch possible future research. 

\section{Related Work}\label{sec:related}
\frage{check how to handle the big data paper with editor}

\frage{todo: Vortex?}

This work is based on two previous papers. In the first,
\cite{dees2013} we consider the queries of the TPC-H benchmark on
a single shared memory machine. We adopt the approach used there to
consider queries manually translated into a single function consisting
of optimized C code. This not only led to performance one or two
orders of magnitude higher than the state of the art but is also
allows us to focus on the algorithmic issues of how to achieve high
performance on modern architectures. The wide adoption of just-in-time
compilation in the mean time further justifies this approach (see
below). Another observation in this paper was that details of
parallelization were orthogonal to the details of how to achieve good
inner loop performance. This motivated us to use the same approach for
evaluating the algorithmic aspects of parallelization on distributed
memory machines. This led to a short conference paper \cite{WDS13}
demonstrating our approach for 6 out of 21 TPC-H queries. The present
full paper extends this to 11 queries, significantly improves the
implementation of 2 of the 3 queries that did not scale well
previously, briefly discusses the remaining 10 queries, describes
additional parallelization techniques, and explains everything in more
detail.

Automatic just-in-time compilation has now become a standard technique \cite{krikellas2010generating,neumann2011efficiently,nagel2014code,freedman2014compilation}
also used by other database systems (e.g.,\ HyPer \cite{kemper2011hyper,rodiger2015high}).
Indeed, our group has also implemented a query-compiler on its own which is part of a commercial product however without a published description of the details. Compile times of such compilers can be in the range of centiseconds with code performance similar to the manually written code we study here.

Early work on cluster query execution for OLAP queries stems from DeWitt et al.\
\cite{dewitt1990parallel} analyzing parallel database systems in a shared
nothing environment running on disk without multi-threading having very low
performance numbers compared to today's systems.
More recent work is from Akal et al.\ \cite{akal2002olap} implementing a database cluster by introducing a middleware for coordination of single cluster nodes without a deep integration into the database system itself.
Their throughput experiments on 64 nodes have similar query times than our implementation, but with a factor of 10\,000 less data.
The work of Akal is further refined by Lima et al. \cite{lima2004adaptive,lima2004olap} by improving load balancing between the nodes and sometimes using indexes instead of scans.
Still, the overall performance problem remains and using data replication makes it difficult to scale to large data sets.

The mentioned solutions do not apply more advanced communication patterns, but it has been shown by Chockler et al.\ \cite{chockler2001group} that this is required to leverage the full performance of the system.
Eavis et al.\ \cite{eavis2010parallel} developed the prototype Sidera which is based on message passing and targets online analytical processing.
Neither the synthetic data, nor the benchmark queries were specified in detail, but performance numbers are in the range of seconds for processing an input data set of 1 million rows on 16 nodes.
Again, our system achieves similar runtimes with the same number of nodes processing about 10\,000 times more data using complicated TPC-H benchmark queries.

Shute et al.~\cite{Shute2013F1} introduce F1, a highly scalable distributed
database for OLTP and OLAP queries which is mainly used for Google AdWords. They
state that their database holds 100\unit{TB} of data and processes hundreds of
thousands of requests per second which is a factor of 3 to 4 more data with
significantly lower runtimes than we achieve in our experiments. However, they do not give additional information on the kind of queries. For large distributed queries they report times similar to MySQL and often linear speedup when adding more resources. While we aim to reduce the use of the network by utilizing copartitioning of tables, they cannot make use of copartitioned tables and cause high network traffic by frequent repartitioning. Our approach becomes therefore more suitable for a scenario with many small machines.

Lee et al.~\cite{Lee2013HANA} give an overview on distributed processing in the SAP HANA database. They also aim to reduce communication and use a toolset that checks incoming workloads and proposes partitioning schemes for the tables. They give experimental results for a database of $100$\unit{GB} with runtimes slightly faster than ours but on a factor 300 less data than we use. They do not give a detailed description of the query used but state that it requires a full table scan over the fact table, which might be comparable to query 1 of the \mbox{TPC-H} benchmark. This is one of the least complex queries we examine and we also achieve fast runtimes for more complex queries.

Cuzzocrea et al.~\cite{cuzzocrea2013olap} propose a framework for parallel building of OLAP data cubes in a shared nothing environment. They evaluate their work on a transformed version of the TPC-H benchmark. Their results show runtimes about a factor 10 faster than in our experiments (excluding the time for building the data cube) but on a factor $3\,000$ less data. Also, their runtimes for some queries are significantly higher (for example query 11, where they report runtimes in the range of minutes instead of centiseconds).

For the case that only the largest tables of the database (fact tables) are split and distributed over the nodes in the cluster and all other tables (dimension tables) are replicated on every node, Furtado et al.~\cite{furtado2005physical} used virtual partitioning to improve load balancing. They replicate the partitions of the fact tables over some nodes and use an algorithm called \emph{adaptive virtual partitioning} to split the tables into virtual partitions that are used to execute parts of the query using a middleware. They evaluate their work on the \mbox{TPC-H} benchmark and show that it is competitive with a full replication of all tables. Because they only provide relative scaling experiments and a relative comparison with the the fully replicated case but no absolute running time, we cannot give a quantitative comparison to our results. However, it is clear that their approach needs significantly more memory than our solution due to replication of the dimension tables and partitions of the fact tables. Also, they only evaluate their implementation on a $SF=5$ \mbox{TPC-H} database which is a factor $6\,000$ less then our experiments.

Han et al.~\cite{han2007progressive} and Karanasos et al.~\cite{karanasos2014dynamically}
both present their approaches to query optimization for distributed query
execution by re-optimizing during execution using accurate statistic
information about the data at the current stage of query execution. As we
optimize query execution by hand, our work is complementary to
theirs, providing insights in possible choices for the optimizer. In particular
Karanasos et al.~\cite{karanasos2014dynamically} show that collecting
information such as the selectivity of predicates before query optimization only
causes minor overhead. This accurate information is required to determine which
of the strategies we use in this paper is most efficient.
\section{Distributed Query Execution}
\label{sec:distributed-execution}
In this section we present our approach to distributed query execution in a shared-nothing environment where all nodes are identical and none plays a special role. We consider these assumptions important for scalability.
The next section describes the data distribution of our system.
After that we present several classes for systematic and efficient query execution in the distributed environment.
Finally, several optimization examples follow.

\subsection{Data Distribution}
In many cases, for the distributed execution of OLAP queries, only the fact tables get partitioned across the nodes, while the dimension tables get replicated across all nodes.
This has the benefit, that most joins between tables can be evaluated locally, eliminating most of the challenges we had to overcome.
The main disadvantage, however, is that these solutions cannot scale well due to
significant memory consumption by the replicated tables.
To support full scalability and large database sizes, we need to minimize replicated data and maximize the usage of the available main memory.
In general, we distribute all tables by partitioning them across the nodes.
Only in extreme cases where a table has a small constant size, we replicate the table across all nodes.
As a result, each node holds $\nicefrac{1}{P}$ of the tuples of each distributed table, where $P$ is the number of nodes.
There are three basic partitioning strategies: range-based, round-robin and hashing \cite{dewitt1992}.
We use range-based partitioning, which is sufficient for synthetic data like in the TPC-H benchmark and also simplifies data generation.
We also use co-partitioning \cite{gao2005consistent}, i.e., for two tables with closely related tuples defined by a foreign key relation, we store corresponding tuples in partitions on the same node.
With this, equi-joins on the foreign key relation can be evaluated locally and additional communication is avoided. 
In our experiments with the database benchmark TPC-H, for example, we use co-partitioning for the tables lineitem and orders and for part and partsupp. \mycomment{\review{ps: TPC-H ist  hier noch nicht eingeführt!}\antwort{m: added forward reference.}}
See a detailed TPC-H table schema in Fig.~\ref{fig:tpch_tables}.
The schema is extended by a data locality property for foreign key relations.
Dashed edges show remote access joins and indicate that the joined tuples can be located in a different partition.
Joins on solid edges can be performed locally.

\usetikzlibrary{positioning}
\begin{figure}
  \centering  \tikzexternaldisable
  \begin{tikzpicture}[->, node distance=2.42cm, state/.style={    rectangle,    draw=black, thick,    minimum height=2em,    inner sep=2pt,    text centered,    align=center  }]
    \node[state] (ORDER) {Orders\\\scriptsize{$SF*1.5\unit{M}$} };
    \node[state, below of=ORDER, anchor=north east, xshift=-.25cm, yshift=1.7cm] (CUST) {Customer\\\scriptsize{$SF*0.15\unit{M}$} };
    \node[state, below of=ORDER, anchor=north west, xshift=.35cm, yshift=1.7cm] (LINE) { Lineitem\\\scriptsize{$SF*6\unit{M}$} };
    \node[state, below of=LINE, yshift=1.2cm] (NATION) { Nation\\\scriptsize{$25$ (replicated)} };
    \node[state, left of=NATION] (REGION) { Region\\\scriptsize{$5$ (replicated)} };
    \node[state, right of=LINE] (PARTSUPP) { Partsupp\\\scriptsize{$SF*0.8\unit{M}$} };
    \node[state, above of=PARTSUPP, yshift=-1.4cm] (PART) { Part\\\scriptsize{$SF*0.2\unit{M}$} };
    \node[state, below of=PARTSUPP, yshift=1.2cm] (SUPP) { Supplier\\\scriptsize{$SF*0.01\unit{M}$} };

    \path[draw, dashed] (CUST.north) |- (ORDER.west);
    \path[draw, line width=1.2pt] (ORDER.east) -| (LINE.north);
    \path[line width=1.2pt] (PART.south) edge (PARTSUPP.north) ;
    \path[draw, line width=1.2pt] (NATION.north) |- +(0.0, 0.3) -| (CUST.south);
    \path[line width=1.2pt] (REGION.east) edge (NATION.west) ;
    \path[line width=1.2pt] (NATION.east) edge (SUPP.west);
    \path[dashed] (SUPP.north) edge (PARTSUPP.south);
    \path[dashed] (PARTSUPP.west) edge (LINE.east);

    \node[draw=none, below of = REGION, xshift=-.8cm, yshift=1.75cm] (ONE) {};
    \node[draw=none, right of=ONE, xshift=-1.8cm] (MANY) {};
    \node[align=left, right of=MANY, xshift=-.5cm] {\footnotesize{Local access (one-to-many)}};
    \path[line width=1.2pt] (ONE) edge (MANY);

    \node[draw=none, below of = ONE, yshift=2.2cm] (ONEREMOTE) {};
    \node[draw=none, right of=ONEREMOTE, xshift=-1.8cm] (MANYREMOTE) {};
    \node[align=left, right of=MANYREMOTE, xshift=-.35cm] {\footnotesize{Remote access (one-to-many)}};
    \path[dashed] (ONEREMOTE) edge (MANYREMOTE);
  \end{tikzpicture}
  \caption{Database tables specified by TPC-H}
  \label{fig:tpch_tables}
\end{figure}
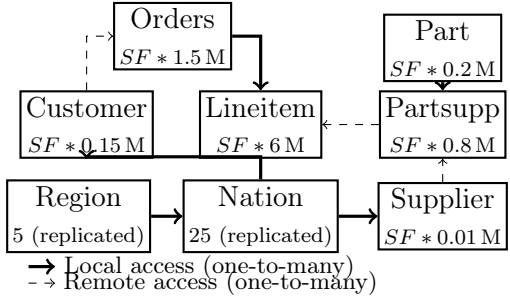

In general, range partitioning can lead to load imbalances, e.g., if a filter predicate qualifies only tuples within a single range of the partitioning key.
In a productive system we would, therefore, rather use hash based partitioning to obtain a reasonable load balance for arbitrary access patterns.
With small modifications, the concepts used here also apply to other partitioning strategies.
\subsection{Efficient Solutions for Data Exchange}\label{sec:efficient_solutions}
\review{ps commented out paragraph intro that was not only grammatically wrong but also void of information.} \antwort{m: ok.}

We now present ways to reduce the communication effort, especially for remote join paths. After a brief discussion of general techniques, we develop more specialized solutions, e.g., by improving filters on remote attributes, and by exchanging bit-reduced, estimated values.

\subsubsection{Data Compression for Reducing Communication Volume}
For the queries we often have to exchange sets of integers (e.g., primary keys, or positions in dictionaries), or, equivalently, very sparse bitsets. Both can be represented by an increasing sequence of integers. These can be compressed by encoding only the differences between subsequent numbers (delta encoding). Various variable-length codes are available for compressing the differences. On the high bandwidth networks we use, we face a trade-off between computation and communication overhead that requires careful (shared-memory parallel and vectorized) implementation of the codecs. We use the  FastPFor library \cite{LemBoy15}, which offers a good compromise in that respect. For unsorted data, dictionary based compression is more effective. Here, the LZ4\footnote{\url{github.com/Cyan4973/lz4}} library gives a good trade-off between speed and compression.

\subsubsection{Filter on Remote Attribute}\label{sec:filter_remote}
Consider the case that the query graph contains a remote join path and the referenced remote attribute is filtered (e.g. ``WHERE $x$.nation=$[nation]$'', with $x$ as a remote relation).
In particular, the (remote) join partners that are qualified by the filter
predicate should be determined. If no column of the remote join partner is used
for the output, this is called a \emph{semi-join}.
We use two different solutions for this problem depending on the table sizes and
the selectivity of the filter.

Alternative~1 is to collect all keys required by the join after all locally
evaluable filters are processed and request them from the remote node. As all
local work has been done, this is the latest point in time possible. Evaluating
local predicates before performing joins has also be found to be beneficial in
other work (e.g. Karanasos et al. \cite{karanasos2014dynamically}). The receiving nodes select qualified rows for the join partner and return a bitset answering for each requested key the question whether the corresponding row qualifies with respect to the filter or not. Using this solution, the amount of additional space required on each node for the filter is independent from the overall size of the table.
Both the sets of requested keys and the reply can be compressed.

Alternative~2 is to filter the remote attribute (join column) and materializing the results as a bitset. Afterwards, the complete bitset is replicated over all nodes (e.g., using the MPI operation all\_gather). Once more, this bitset can be compressed. With this, we avoid the explicit transmission of the required keys. We can profit in cases where most nodes address a significant fraction of the remote table anyway or when the remote filter is highly selective. In that latter case, an additional benefit is that local work can be reduced by applying the remote filter first which is just an access to the replicated bitset.

In a productive system, the choice between these two alternatives can be made by
estimating the selectivity of local and remote filters using sampling. Using the
known table sizes, the number of nodes, and appropriate models for the cost of
the collective communication operations \cite{bruck1997efficient}, one can then
approximate the overall cost of both alternatives. Karanasos 
et al. \cite{karanasos2014dynamically} show that performing a \emph{pilot run} to collect these information only has little
overhead.
To make this more concrete, we estimate the number of bits communicated by each node assuming random distribution of data, information theoretically optimal compression, $P$ nodes, $n$ requests generated after local filtering ($n/P$ per node),
a remote table of size $m$ and $\gamma m$ rows of the remote table surviving remote filtering. Alternative~1 then requires $\nicefrac{n}{P}\log\nicefrac{mp}{n}$ bits of communication%
.\footnote{This expression only makes sense if $n/p<m$. However for $n/p>m$, Alternative~2 is better anyway.}
Alternative~2 communicates $\gamma m\log\nicefrac{1}{\gamma}$ bits.

\subsubsection{Selecting the Global top-$k$ Results From Local Ones}\label{sss:globaltopk}
A prominent pattern of decision support queries is aggregating values by key and returning only the top-$k$ results.
Assuming that the data is partitioned by the key used for aggregation,
we can aggregate locally and then identify the global top-$k$ elements among the results.
This is the classical selection problem in a distributed setting.
Asymptotically efficient algorithms have been considered in the literature (e.g. \cite{HubSan16}). Here we consider simple, pragmatic solutions. First, it makes sense to identify the local top-$k$ results on each processor.
A na\"{\i}ve solution would then be to gather all  $P \cdot k$ results on one
root node, sort them and only keep the first $k$ rows. By making use of the
collective \emph{reduce} operation, we can reduce the communication effort: 
The input to the reduction are the $k$-vectors of locally largest values sorted in descending order.
Every time the messages of two nodes get combined by the reduce function, we merge the two (sorted) arrays and only keep the first $k$ rows. Since the messages sizes for both solutions are equal, the bottleneck communication volume for the reduce operation is logarithmic in the number of MPI processes in contrast to linear for the gather operation, we take some load off the network with this approach.

\subsubsection{Filtering top-$k$ results} \label{sec:top-k-filter}

Now consider a similar situation as in Section~\ref{sss:globaltopk}. Aggregation is possible locally but now
some of the keys are disqualified, e.g. by a filter condition.
The interesting case is when the filter qualifying or disqualifying the keys lies on a remote join path.
We use an algorithm that reduces communication overhead by evaluating the remote filter in a lazy fashion.
We request the filter results only for chunks of so far unfiltered elements that have locally largest values.  Assuming that a fraction $p$ of the keys qualify for the result, we only need to communicate data for expected $\nicefrac{k}{p}$ keys instead of all the keys.
This iteration ends when each PE has identified $k$ elements that pass the filter condition. Then the globally best elements are determined as in Section~\ref{sss:globaltopk}. If $p$ is very small, one can optionally run a global top-$k$ identification from time to time. Only PEs that still have unfiltered elements larger than the globally $k$-largest filtered elements then need to continue filtering.


\subsubsection{Top-$k$ Selection on Distributed Results}
\label{sec:topk}
A more difficult case is when the values to be aggregated are
not partitioned by key.
The complete aggregate of each key is found by aggregating the partial results from all nodes.
One na\"{\i}ve solution for this problem is to compute all complete aggregate results from the partial results and determine the top-$k$ results afterwards.
However, in the case of many keys and small $k$, the communication overhead for this operation can be very high compared to the final result size.
There has been a lot of previous work for solving this problem efficiently, for example, the threshold algorithm by Fagin et al.\ \cite{fagin2001optimal} or TPUT by Cao et al.\ \cite{cao2004efficient}.
Unfortunately, these algorithms do not perform well with the aggregation function \texttt{SUM} if we have the same independent value distribution of the partial sums across the nodes.
In this case, the final aggregated sums follow a normal distribution and both algorithms communicate almost all partial sums before selecting the top-$k$.

For this situation we propose a new distributed algorithm that communicates only several bits of all partial sums.
Full values are only communicated for a small set of top-$k$ candidates.
A detailed description of the algorithm follows.

In the first step, we approximate each partial sum by only $m$ bits of the number.
To skip leading zeros, the $m$ bits begin at an offset which is shared by a group of keys (e.g.,\ $1024$).
The offset is equal to the position of the highest one-bit of all numbers within the group.
These $m$ bits are only an approximation of the values as lower bits are missing. Still, we can compute a maximal and minimal error (all lower bits are one and zero, respectively).
Each node is now responsible for a range of keys, which are distributed by a personalized all-to-all message such that each node receives all encoded sums for its key-range.
We further compute a lower and upper bound for each decoded partial sum and sum them up by key, resulting in an upper and lower bound for the total sum per key.
A collective reduce operation determines the global k-th highest lower bound.
Each key with an upper bound below the k-th highest lower bound cannot be part of the top-$k$ results anymore and is, therefore, discarded.
After that, each node requests the full partial sums for its remaining keys, which is expected to be a small set.
In a final step, the k-th highest total sums are determined across the nodes.

Using a larger number for the number of bits $m$ increases the message size in the first step but also improves the lower and upper bound afterwards.
In our experiments, we applied this algorithm in query 15 of the TPC-H benchmark reducing the communicated data volume by a factor of $8$ compared to the na\"{\i}ve solution (see Section~\ref{sec:exp-top-k}).

\review{ps: although I cannot follow you description my impression is that this approach is not robust to arbitrary inputs and measurements with the random data from TPC-H is unfair. Why not use robust sampling based techniques?}\antwort{m: man uebertraegt im w.c. nie mehr (bis auf offsets) als alle werte einzeln. Bei normal/gleichverteilung ist jedoch die selektivitaet viel hoeher und es bleiben nur noch wenige kandidaten uebrig.}\antwort{m: The approach should be robust. One can approach normally-distributed values for practical applications; therefore the final candidates set should be small. In the worst case, all values are equal -- our solution is as robust as the trivial solution. sampling would not help here because all samples would be equal but one needs to check anyway if all values are equal by aggregating all of them. Nevertheless, I see a problem regarding an increasing bias due to larger $P$. More encoding bits would solve this.}

\subsubsection{Tuning Basic Communication Functions} \label{sec:tuning-basic-communication-functions}

For communication between the nodes, we use collective operations provided by our MPI implementation.
Here, operations like all-to-all, gather or reduction are implemented in an efficient and usually non-trivial way.
But even a dedicated framework like MPI can suffer performance problems at certain functions, which we noticed during experiments for all-to-all in our OpenMPI library v1.8.4 implementation.
The average all-to-all throughput of sent data per node changed from $0.5$\unitfrac{GB}{s} to $2.5$\unitfrac{GB}{s} when switching from 12 nodes to 16 nodes. Also, we noticed a high variance between different runs.

To tackle the performance problem, we use our own implementation of a
personalized all-to-all communication using the 1-factor algorithm
\cite{sanders2002:1factor}.  It uses non-blocking send and receive
calls for point-to-point messages exchange.  The algorithm requires
$\mathcal{O}(P)$ communication rounds for pairing each node with each
other node and is thus linear in the number of nodes.  A communication
partner of $u$ in round $i$ is $v^i(u)=(i-u)\bmod p$.\footnote{Note
  that $u$ is also the partner of $v^i(u)$, which can be seen by
  evaluating $v^i(v^i(u))=(i-((i-u)\bmod p))\bmod p = u$} The 1-factor
algorithm is faster compared to the library-provided all-to-all by at
least a factor of two in our micro-benchmark.

\subsubsection{Late Materialization}

Analytical query results often consist only of a small number of rows as the answer should remain human-readable.
Actually, this is true for all 22 query results of the TPC-H benchmark and usually achieved by small group-by cardinalities or selecting only the top $k$ results.
Consequently, we delay the gathering of secondary attributes in the result set that are not involved in the actual query computation (e.g.\ in TPC-H query 15: s\_name, s\_address, s\_phone).
This way, the secondary attributes do not slow down the main query computation. 
When the final result is collected on a single node, we can request the attributes by one collective scatter operator and receive them by a collective gather operation both in $\mathcal{O}(\log P)$ steps, where $P$ is the number of processors. 




\section{Application in TPC-H}
\label{sec:applicationtpch}
\review{ps todo: describe TPC-H similar to ICDE paper. Also give abstract description of implemented queries.}
\subsection{TPC-H}
The TPC-H benchmark is used to measure the performance of database systems for decision support (OLAP) queries \cite{council2008tpc,Poess:tpch}.
We use the data generator defined by the benchmark and check the query results for correctness.
We do not change the ordering of the rows in the tables.
Each table is split into $P$ (number of nodes) chunks and chunk $i$ is generated directly in main memory on node with rank $i$ using the following dbgen parameters: -s $\langle SF\rangle$ -S $\langle rank\rangle$ -C $\langle P \rangle$.
Only the tables NATION and REGION with both at most 25 rows are not split and replicated across all nodes.
We implement 11 out of 22 TPC-H queries covering several aspects like filtering, small and large aggregations and different join types.
Section~\ref{sec:the-implemented-queries} gives a detailed description of the queries.


To allow fair comparison with other systems, we comply with the official TPC-H rules as far as possible.
In particular, we follow the rules for sorting relations, data structures, and join indexes, which are created transparently between all foreign keys.
Still, we do not provide the full functionality of a DBMS: We do not support ACID, updates, and the execution of arbitrary SQL statements.
See the discussion on future work in Section~\ref{sec:conclusion} for more details.



\subsection{Parallelization}
We use a hybrid parallelization approach for the implementation combining inter-node and intra-node parallelism.
For the inter-node parallelism we use the open standard MPI (message passing interface), which provides collective communication operations for remote data exchange \cite{gropp1999using}.
Our MPI implementation is Open MPI \cite{gabriel2004open}, an open source implementation of the MPI specification. \mycomment{jd, removed: is applied for the coordination and synchronization.}
The collective operations used by our algorithms are gather (collecting a message from each node at root), allgather (like gather but every node gets the messages from all nodes), scatter (send a message to each node from root), all-to-all (every node exchanges a message with every node), reduce (every node has a message, all messages are the same size, and an operator is applied when joining two messages, the result lies on one root node) and allreduce (like reduce but every node gets the result).
Moreover, we implement user-defined reduce operators for an efficient result aggregation as well as customized MPI data types.

Besides MPI, intra-node parallelism based on shared-memory is realized by using OpenMP for simple loop parallelization and TBB (Intel Threading Building Blocks), a template library for C++ that offers an abstraction of thread management \cite{reinders2010intel}.\review{ps: somewhere discuss NUMA issues?}
In general we apply data-parallelism and logically partition the input into several parts for processing using ``parallel\_for'' and ``parallel\_reduce'' of the TBB framework, providing work stealing and load balancing between the threads.
This way we take full advantage of the available intra-node parallelism.

\subsection{The Implemented Queries}
\label{sec:the-implemented-queries}
We select queries 1, 2, 3, 4, 5, 11, 13, 14, 15, 18 and 21 from the
$22$ TPC-H queries with the objective to cover various challenges and
access patterns for distributed
execution. Appendix~\ref{app:notImplemented} discusses the remaining
queries, indicating that only few of them would raise additional
questions.  There is also a certain focus on expensive queries.

Query 1 performs a large aggregation and accesses only a single table, providing the top ten unshipped orders based on the potential revenue per order.
It is the most used query in related work.

Query 4 refers to two co-partitioned tables.
It counts per order priority ($5$ distinct values) the number of orders, which contain delayed lineitems to estimate the quality of the order priority system. 

Query 18 also uses two co-partitioned tables. It only accesses remote attributes for the result output.
It determines the top-$100$ customers based on the property of having placed a large quantity order.

The remaining queries have significant remote data dependencies, which means that join partners can be stored on a different partition.

Query 2 uses none of the fact tables but has a remote filter attribute to determine qualified suppliers and get the top-100 results.

Query 3 uses two fact tables and one remote attribute as filter to provide the top ten unshipped orders based on the potential revenue per order.

Query 5 uses one fact table and two filter attributes on remote join paths. The result consists of only five rows.

Query 11 uses no fact table and has no locally evaluable filter. It has a filter on a remote attribute and a threshold filter that is dependent on a global aggregation.

Query 13 uses one fact table and a filter attribute on a remote join path. It also groups by a key on that remote join path.

Query 14 uses one fact table and a remote filter attribute. The result consists of only one row and is computable using two aggregates.

Query 15 uses one fact table and remote attributes for result output.
It produces a large intermediate set of partial results (grouped by a remote key) where we want to find the top-1 element only.

Query 21 is similar to query 15 but additionally applies a remote filter during aggregation.

We continue with a detailed implementation description for each query.
Note that we perform local aggregations of a query using shared-memory parallelism (as described in \cite{dees2013} where applicable) and we do not mention them explicitly. We encourage the reader to check the SQL code for the queries from the \mbox{TPC-H} specification in order to follow the detailed descriptions.

\emph{Query 1} (pricing summary report) reports the overall amount of business that was billed, shipped and returned within a time interval.
At first, the query aggregates a key figure based on the lineitem table.
The aggregates are grouped by two possible \emph{returnflags} and three different values of \emph{linestatus}.
Therefore, the distributed result set has $6$ entries at most.
Second, we use the collective reduce operation to aggregate the distributed results.
A custom reduce operator merges the partial result sets by \emph{returnflags} and \emph{linestatus}.

\emph{Query 2} (minimum cost supplier) finds for each part of a given size and type the
supplier from a given region with the lowest price for that part and returns the top-100 results ordered by the suppliers account balances. After filtering by size and type, only 0.4\% of the partsupps remain to be filtered by the suppliers region, so we request these filter results explicitly. After we found all suppliers that qualify for the result, we send this information to the corresponding nodes, sort the suppliers by their account balance and derive a global result using a custom reduce function. As the query uses some columns that are not required for computing the top-100 results, we can safe a significant amount of communication time by materializing these columns at the latest point in time possible, which is when the global top-100 results are found.

\emph{Query 3} (shipping priority) provides the top ten unshipped orders based on the potential revenue per order. We implement two versions for this query.
For the first version, we transform the query into two sub-queries to resolve remote dependencies.
The first sub-query computes an intermediate result by applying the second solution from Section~\ref{sec:filter_remote}, where a filter is evaluated on a join attribute to qualify customers by their nation. 
Afterwards, the intermediate results are redistributed.
The second sub-query uses the intermediate result to filter and aggregate. 
Thus, it operates on locally available data.
Finally, each node keeps the local top-ten result tuples.\\
For the second version, we use the solution from Section~\ref{sec:top-k-filter}. We first aggregate and filter with the locally available information. We then sort the orders by their revenue and request the filter result on the customers market segment until we have found the top ten results on each node.\\
A collective reduce operation gains the global top-ten in both versions. In particular, we implement a custom reduce operator that selects the top-ten of two incoming local top-ten lists.

\emph{Query 4} (order priority checking) counts per order priority ($5$ distinct values) the number of orders, which contain delayed lineitems to estimate the quality of the order priority system.
The lineitems of a qualified order are aggregated by the corresponding priority.
The distributed results are aggregated using a collective reduction.

\emph{Query 5} (local supplier volume) lists the revenue done through customers and suppliers from the same nation during the period of one year in a given region. Due to the small size of the supplier table, we distribute their nation over all nodes. We then filter the orders by year and the suppliers region and request the nations for all required customers. After receiving the nations, we filter by the customers nation, group the orders and derive a global result by a collective reduce operation.

\emph{Query 11} (important stock identification) reports the parts that are (in terms of value) most available in a given nation. Because there is no locally evaluable filter, we distribute the filter result on the suppliers nation over all nodes. We then calculate the total value of all available parts in the given nation locally and derive the global sum using the allreduce operation. After that we can select all qualified parts and gather them on one node.

\emph{Query 13} (customer distribution) reports how many customers have placed 1, 2, 3, $\dots$ orders. Only orders matching a filter condition qualify. We split this query into two sub-queries. First, we get all customers of qualified orders and send their keys to the corresponding nodes. We then compute a local result and derive the global result.

\emph{Query 14} (promotion effect) calculates the fraction of revenue done by special parts over all revenues during one month. We split this query into two subqueries. First, we filter the lineitems by date. We then request the filter on the remote join path to the parts and calculate the total revenue as well as the revenue done through promotion parts. After that, we reduce the global and the promotion revenue to the root node in order to calculate the final result.

\emph{Query 15} (top supplier) determines the suppliers with a maximum total revenue based on lineitems within a specified time interval.
We split the original query into two sub-queries for resolving the remote dependencies.
The first sub-query aggregates the revenue per supplier. The join path to supplier is remote.
Hence, every node has knowledge only of the partial revenue per supplier.
The output should contain the maximum revenue and the related suppliers.
Consequently, we apply our top-$k$ selection algorithm with value approximation (see Section~\ref{sec:topk}) to determine the maximum total revenue. 
Alternatively, we also implemented the simple solution of redistributing all partial sums to their corresponding nodes (determined by their partition key), aggregate them and select the maximum.
We expect a better performance of the approximation algorithm over the simple solution.

\emph{Query 18} (large volume customer) determines the top-$100$ customers based on the property of having placed a large quantity order.
The query aggregates lineitems and reports the top-$100$. In a first step, the local top-$100$ are determined.
Afterwards, the local results are reduced to select the global top-$100$.
The result output contains attributes of remote join paths.
Therefore, we request the remote attributes for the $100$ tuples and collect them.

\emph{Query 21} (suppliers who kept orders waiting) lists those
suppliers of a specified nation who were part of a multisupplier order
and were the only supplier delaying the order. We implement two
versions for this query.  For the first version we transform the query
into three sub-queries.  The first sub-query computes an intermediate
result by evaluating a filter on the join attribute (see the second
solution from Section~\ref{sec:filter_remote}) qualifying suppliers by
their nation.  The intermediate results are redistributed and used in
the second sub-query to filter tuples during the aggregation.  In
particular, the aggregate is grouped by a remote attribute, which
implies a distributed result among the nodes.  The partial results are
aggregated within a third sub-query to select the final top-ten
tuples.\\ 
For the second version we compute the intermediate result
without the filter on the suppliers nation. We then request the filter
result for the suppliers nation for all suppliers 
that hold up at
least one delayed shipment (see the first solution from
Section~\ref{sec:filter_remote}).
For every qualified supplier, the
number of delayed shipments is then gathered at their corresponding
nodes.\\ The local top 100 suppliers are kept and the global result is
determined using a collective reduce operation.

\mycomment{
Here, we present a pseudo-code for each of the implemented six TPC-H queries (Procedure~\ref{algo:01}--Procedure~\ref{algo:21}).\review{ps: Der Pseudocode ist für mich unverständlich. Entweder Notation viel besser erklären oder (besser) auf verbale Beschreibung umstellen. } \antwort{m: siehe naechste anmerkung}
We only listed code parts which are important for the clustered execution.
Moreover, we added an annotation below the pseudo-code, if the query was split up into sub-queries.
Under this circumstance, we also indicated the applied pattern to resolve the remote dependency.
Additionally, the shared-memory aggregation algorithms are similar to those listed in \cite{dees2013} and in these terms, we did not list details.
}
\review{ps query Beschreibungen wieder einschalten? Oder sind die jetzt woanders. Ganz ohne ist nicht akzeptabel.} \antwort{m: Vllt. den pseudocode komplett entfernen und dann textuell je query: 1 satz was die query macht, dann 1-2 saetze wie geloest. Dann spart man sich den pseudocode auch.}

\mycomment{Before the pseudo-code, each query is set into a semantic context. \par
Query 1 (pricing summary report) reports the overall amount of business which was billed, shipped and returned.
See the pseudo-code in Procedure~\ref{algo:01}.
Query 3 (shipping priority) provides the top ten unshipped orders based on the potential revenue per order.
Moreover, the associated customer's marketsegment of the order, the order date and the date of the related lineitems are filters.
Procedure~\ref{algo:03} illustrates the query execution.
Query 4 (order priority checking) counts per order priority ($5$) the number of orders, which contain delayed lineitems to estimate the quality of the order priority system.
See Procedure~\ref{algo:04} for the pseudo-code.
Query 15 (top supplier) determines the suppliers with a total maximum revenue based on lineitems within a specified time interval.
The pseudo-code is shown in Procedure~\ref{algo:15}.
Query 18 (large volume customer) determines the top $100$ customers based on the property of having placed a large quantity order.
Procedure~\ref{algo:18} contains the pseudo-code.
Query 21 (suppliers who kept orders waiting) lists those suppliers of a specified nation, who were part of a multisupplier order and they were the only supplier who caused a delay.
In terms of the results, only the top $100$ are returned, sorted by their number of delayed orders.
See Procedure~\ref{algo:21} for the pseudo-code.}

\newcommand{\MPI}[1]{\STATE \textbf{mpi} #1}
\renewcommand*{\algorithmcfname}{Procedure}

\review{ps:I get undefined control sequence errors in latex here.}\antwort{m: no pseudocode, everything textual now.}
\review{ps: ich kann den Kommentaren von m nur zustimmen. Wenn man es drinlaesst sollte man das Aus dem TPC-H Kapitel herausziehen und als allgemeines Ergebnis erklären.}\antwort{m: ich denke es kann komplett raus, da es zwar eine optimierung ist, aber sehr technisch; zudem hat es mit den hauptcontributions nicht soviel zu tun dass man es erklaeren muss; weiterhin platz knapp}

\section{Evaluation}
In this section, we evaluate the combination of a clustered query execution using message passing for the inter-node communication, with shared-memory parallelism on each node and highly optimized algorithms.
In this context, all tables (except extremely small tables with $\leq 50$ rows) are range-partitioned without table replication.
For query 3 and 21, we also evaluate the behavior if the remote join attribute is replicated. 
\subsection{Methodology}
\review{ps: diese Einleitung ist ziemlich länglich. Im übrigen frage ich mich ob man aus Plaztgründen strong scaling komplett wegläßt bzw. nur erklärt warum das sinnlos ist (Eingabe auf einem Knoten so klein, dass Laufzeiten auf vielen Knoten lächerlich klein werden. weak scaling ist auch aus Anwendungssicht, dass was man haben will, weil wir bei SF 100 ja schon interaktive Laufzeiten für alle queries haben.} \antwort{m: Guter punkt, die reingemischte strong-scale analyse lenkt vllt. auch vom schwerpunkt ab, wir wollen ja clustered query execution machen, dass wir vor allem groessere datenmengen verarbeiten koennen - ich forme dieses chapter entsprechend um.}
We measure the running time and scalability of the implemented queries to evaluate our contribution.
In this context, \emph{weakly scaled} factors are used to linearly scale up the input size with the number of computation nodes \cite{poly3}.
This approach simulates the case of an end user who wants to run distributed queries on a growing database.
The configurations for $\{\text{\#nodes}, \text{scale factor}\}$ were  $\{2^i; 100\cdot 2^i\}$ for $i=0..7$. 
\mycomment{\frage{m: removed} Moreover, we do not present the \emph{strong scaling} results because the input sizes are drastically small with an increasing number of nodes, such that the network communication dominates the running time.}
We briefly introduce the technical method that was used in the implemented prototype for measuring the experiments.
At first, we synchronize the nodes with a barrier before each query run. 
Second, we measured the walltime for the complete query execution.
The walltime is used because communication times are hidden from the local CPU time \cite{bader2002algorithm} but should be considered in the measures.
Third, in order to get an accumulated communication time per query, we also track the running time of occurring MPI communication operations.
In detail, the walltime values of each node were aggregated on the root node to determine the mean running time over all nodes.
Additionally, specific checkpoints were tracked by using the CPU time.
Those detailed measures allow the evaluation of shared-memory parallelism.
\\

\subsection{Experimental Setup}
For our experiments we use a cluster where each node has $64\unit{GB}$ main memory and two E5-2670 Intel Xeon octa-cores with $2.6\unit{GHz}$, $8\times256\unit{KB}$ L2 cache, and $20\unit{MB}$ L3 cache. Up to 128 out of 400 nodes are available per user.
The nodes are connected using InfiniBand 4X QDR.
According to the cluster user manual, the point-to-point network bandwidth is more than $3\,700\unitfrac{MB}{s}$ with a latency about $1\unit{\mu s}$.
We ran micro-benchmarks to measure the real throughput, (a) using explicit send/receive (between two nodes $3\,480 \unitfrac{MB}{s}$) and (b) using a personalized all-to-all (between $2 - 8$ nodes $\approx 3\,000 \unitfrac{MB}{s}$, $P\ge 16:\,  <2\,400\unitfrac{MB}{s}$ in Open MPI v1.8.4)\review{ps: bezüglich welcher Implementierung. }\antwort{m: ich habe name und version der referenzierten implementierung ergaenzt.}.
The experienced throughput is, therefore, lower than promised.
This observation is critical because we use only collective operations for inter-node synchronization, such as all-to-all. 
The cluster (thin nodes) allows a theoretical maximum main-memory usage of $8\unit{TB}$.
A Suse Linux Enterprise (SLES) 11 runs on every node.
We compile our implementation with GCC 4.8.5 (optimization level -O3) and use Open MPI 1.8.4 as message passing library.

\subsection{Experiments}
\review{ps: Queries einheitlich nummerieren mit oder ohne null in 01 etc.}
The results of our first experiment are presented in Fig.~\ref{fig:eval:no_remote_agg}, which contains the plotted running times and Fig.~\ref{fig:eval:communication}, which contains the percentage of time spent for communication, both for weakly scaled factors. Some queries have been tested in several variants.
Label \emph{late} stands for the first method of remote filtering described in Section~\ref{sec:filter_remote} -- request data after local filtering.
Label \emph{repl}(icate) stands for a version where the table on the remote join path is replicated over all nodes, allowing a local evaluation of the join. Label \emph{lazy} refers to the top-$k$ filtering method from Section~\ref{sec:top-k-filter}. The versions of query 3 and 21 without any addition use the second method from Section~\ref{sec:filter_remote}.

Queries 1, 4 and 18 only require data during the aggregation which are available on the node's partition.
In this context, we expected a constant running time in the weak scaling experiment.
As evident from Fig.~\ref{fig:eval:no_remote_agg}, the running times were nearly constant.
The maximum scale factor of the experiment was $12,800$ on $128$ nodes.
In the experiment, the queries 4 and 18 required around $80$--$130\unit{ms}$, whereas query 1 requires $\approx 270\unit{ms}$ for execution.

The main challenge for queries with join paths to tuples on a non-local partition  (queries 2, 3, 5,11, 13, 14, 15, 21) was the reduction of intermediate communications.
Those communications represent an inherent sequential part of the query execution and moreover, the message sizes depend on the scale factor.
Therefore, it is required to keep them small enough to gain good scale up characteristics in order to increase the number of nodes ($P$) for growing hardware and computation demands.
The running times for weakly scaled factors should increase for larger $P$ because of an increased communication effort for joining or redistributing intermediate sub-query results.

As can be seen in Fig.~\ref{fig:eval:no_remote_agg}, the running times of queries 3, 15 and 21 (without replication) increased with $P$ and its corresponding scale factor.
Nevertheless, the running time did not double for a doubled input size and factor two more nodes.
For example, the execution of query 15 took four times longer on $64$ nodes than on one node, although the amount of processed data was $64$ times higher.
The observed increasing running time can be explained with an increasing communication effort since the number of communicated elements doubled for each step on the x-axis.

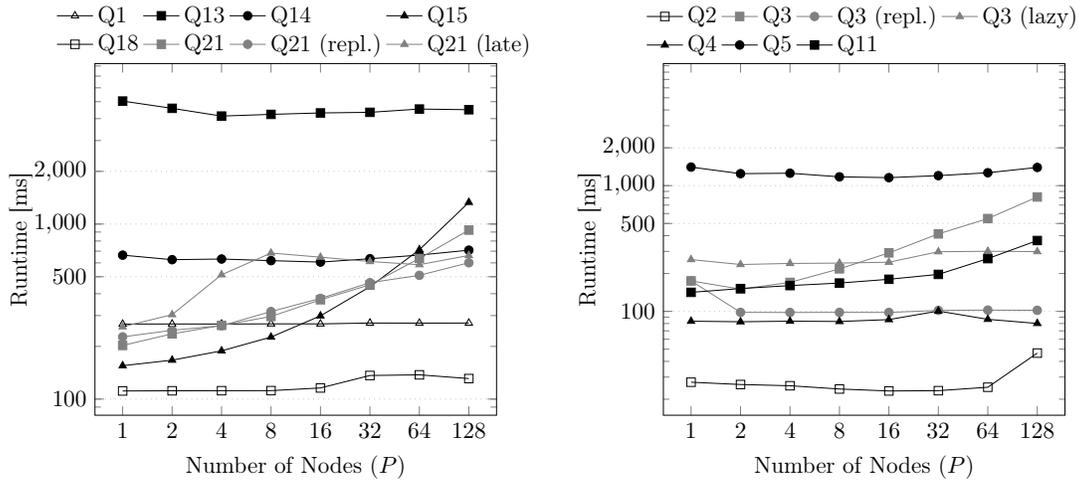
\begin{figure}[ht]
  \centering
    \pgfplotsset{/pgf/number format/.cd,fixed,precision=0} 
    \tikzsetnextfilename{weak-scaling-1}
    \begin{tikzpicture}[scale=.8]
      \begin{loglogaxis}[log basis x=2,  log basis y=10, xmin=1, ymin=0, ymax=6000, extra y ticks = {500, 2000}, enlargelimits=.08,
        xmax=128, xtick=data,
        xlabel={Number of Nodes ($P$)},
        xticklabel=\pgfmathparse{2^\tick}\pgfmathprintnumber{\pgfmathresult},
        yticklabel={\pgfmathparse{10^\tick}\pgfmathprintnumber[fixed]{\pgfmathresult}},
        ylabel={Runtime [ms]}, major grid style={dotted},ymajorgrids = true,
        legend cell align=left,
        legend style={at={(.5, 1.0)}, anchor=south, draw=none},
        legend columns = 4,
        height=0.45\textwidth,
        width=0.5\textwidth,
        ]
        \addplot[color=black, mark=triangle] table [x=procs, y expr={\thisrow{runtime}*0.6+\thisrow{peaktime}*0.4}] {timings/q01.time};
        \addlegendentry{Q1};

        \addplot[color=black,  mark=square*] table [x=procs, y expr={\thisrow{runtime}*0.6+\thisrow{peaktime}*0.4}] {new_timings/Q13_QV1.txt};
        \addlegendentry{Q13};

        \addplot[color=black, mark=*] table [x=procs, y expr={\thisrow{runtime}*0.6+\thisrow{peaktime}*0.4}] {new_timings/Q14_QV1.txt};
        \addlegendentry{Q14};

        \addplot[color=black, mark=triangle*] table [x=procs, y expr={\thisrow{runtime}*0.6+\thisrow{peaktime}*0.4}] {timings/q15-weak.time};
        \addlegendentry{Q15};

        \addplot[color=black, mark=square] table [x=procs, y expr={\thisrow{runtime}*0.6+\thisrow{peaktime}*0.4}] {timings/q18.time};
        \addlegendentry{Q18};

        \addplot[color=gray, mark=square*] table [x=procs, y expr={\thisrow{runtime}*0.6+\thisrow{peaktime}*0.4}] {timings/q21-weak-scale-qv2.time};
        \addlegendentry{Q21};
        \addplot[color=gray, mark=*] table [x=procs, y expr={\thisrow{runtime}*0.6+\thisrow{peaktime}*0.4}] {timings/q21-weak-scale-qv1.time};
        \addlegendentry{Q21 (repl.)};

        \addplot[color=gray, mark=triangle*] table [x=procs, y expr={\thisrow{runtime}*0.6+\thisrow{peaktime}*0.4}] {new_timings/Q21_QV3.txt};
        \addlegendentry{Q21 (late)};

      \end{loglogaxis}
    \end{tikzpicture}
    \pgfplotsset{/pgf/number format/.cd,fixed,precision=0} 
    \tikzsetnextfilename{weak-scaling-2}
    \begin{tikzpicture}[scale=.8]
      \begin{loglogaxis}[log basis x=2,  log basis y=10, xmin=1, ymin=0, ymax=6000, extra y ticks = {500, 2000}, enlargelimits=.08,
        xmax=128, xtick=data,
        xlabel={Number of Nodes ($P$)},
        xticklabel=\pgfmathparse{2^\tick}\pgfmathprintnumber{\pgfmathresult},
        yticklabel={\pgfmathparse{10^\tick}\pgfmathprintnumber[fixed]{\pgfmathresult}},
        ylabel={Runtime [ms]}, major grid style={dotted},ymajorgrids = true,
        legend cell align=left,
        legend style={at={(.5, 1.0)}, anchor=south, draw=none},
        legend columns = 4,
        height=0.45\textwidth,
        width=0.5\textwidth,
        ]

         \addplot[color=black, mark=square] table [x=procs, y expr={\thisrow{runtime}*0.6+\thisrow{peaktime}*0.4}] {new_timings/Q02_QV1.txt};
        \addlegendentry{Q2};

          \addplot[color=gray, mark=square*] table [x=procs, y expr={\thisrow{runtime}*0.6+\thisrow{peaktime}*0.4}] {timings/q03-weak-scale-qv2.time};
        \addlegendentry{Q3};
        \addplot[color=gray, mark=*] table [x=procs, y expr={\thisrow{runtime}*0.6+\thisrow{peaktime}*0.4}] {timings/q03-weak-scale-qv1.time};
        \addlegendentry{Q3 (repl.)};
        \addplot[color=gray, mark=triangle*] table [x=procs, y expr={\thisrow{runtime}*0.6+\thisrow{peaktime}*0.4}] {new_timings/Q03_QV4.txt};
        \addlegendentry{Q3 (lazy)};

        \addplot[color=black, mark=triangle*] table [x=procs, y expr={\thisrow{runtime}*0.6+\thisrow{peaktime}*0.4}] {timings/q04.time};
        \addlegendentry{Q4};

		\addplot[color=black, mark=*] table [x=procs, y expr={\thisrow{runtime}*0.6+\thisrow{peaktime}*0.4}] {new_timings/Q05_QV1.txt};
        \addlegendentry{Q5};

        \addplot[color=black, mark=square*] table [x=procs, y expr={\thisrow{runtime}*0.6+\thisrow{peaktime}*0.4}] {new_timings/Q11_QV2.txt};
        \addlegendentry{Q11};
      \end{loglogaxis}
    \end{tikzpicture}
  \caption{Experimental results ($SF=100\cdot P$) - runtime}
  \label{fig:eval:no_remote_agg}
\end{figure}
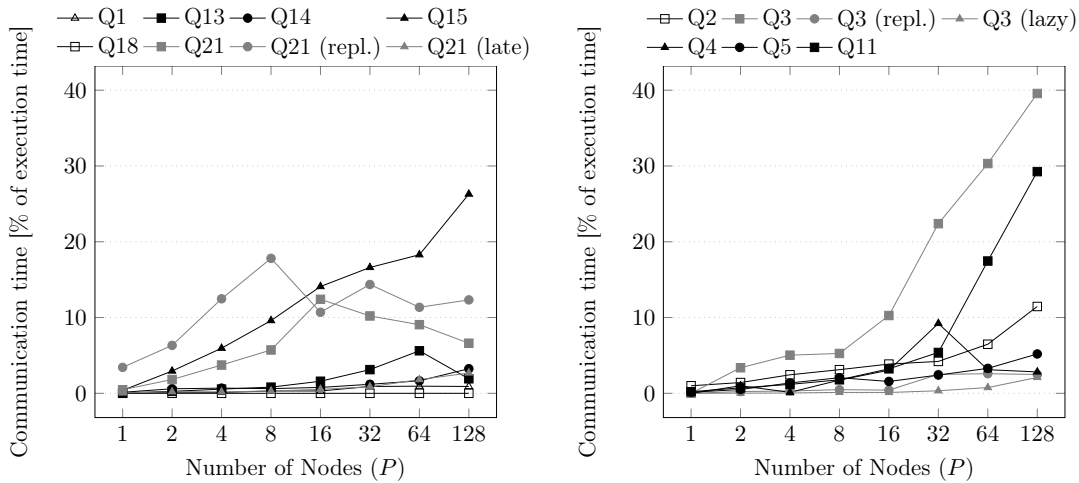
\begin{figure}[h!t]
  \centering
    \pgfplotsset{/pgf/number format/.cd,fixed,precision=0} 
    \tikzsetnextfilename{weak-scaling-comm-1}
    \begin{tikzpicture}[scale=.8]
      \begin{semilogxaxis}[log basis x=2, xmin=1, ymin=0, ymax=40, enlargelimits=.08,
        xmax=128, xtick=data,
        xlabel={Number of Nodes ($P$)},
        xticklabel=\pgfmathparse{2^\tick}\pgfmathprintnumber{\pgfmathresult},
        ylabel={Communication time [\% of execution time]}, major grid style={dotted},ymajorgrids = true,
        legend cell align=left,
        legend style={at={(.5, 1.0)}, anchor=south, draw=none},
        legend columns = 4,
        height=0.45\textwidth,
        width=0.5\textwidth,
        ]
        \addplot[color=black, mark=triangle] table [x=procs, y expr={\thisrow{commpart}}] {timings/q01.time};
        \addlegendentry{Q1};

        \addplot[color=black,  mark=square*] table [x=procs, y expr={\thisrow{commpart}}] {new_timings/Q13_QV1.txt};
        \addlegendentry{Q13};

        \addplot[color=black, mark=*] table [x=procs, y expr={\thisrow{commpart}}] {new_timings/Q14_QV1.txt};
        \addlegendentry{Q14};

        \addplot[color=black, mark=triangle*] table [x=procs, y expr={\thisrow{commpart}}] {timings/q15-weak.time};
        \addlegendentry{Q15};

        \addplot[color=black, mark=square] table [x=procs, y expr={\thisrow{commpart}}] {timings/q18.time};
        \addlegendentry{Q18};

        \addplot[color=gray, mark=square*] table [x=procs, y expr={\thisrow{commpart}}] {timings/q21-weak-scale-qv2.time};
        \addlegendentry{Q21};
        \addplot[color=gray, mark=*] table [x=procs, y expr={\thisrow{commpart}}] {timings/q21-weak-scale-qv1.time};
        \addlegendentry{Q21 (repl.)};

        \addplot[color=gray, mark=triangle*] table [x=procs, y expr={\thisrow{commpart}}] {new_timings/Q21_QV3.txt};
        \addlegendentry{Q21 (late)};

      \end{semilogxaxis}
      
    \end{tikzpicture}
    \pgfplotsset{/pgf/number format/.cd,fixed,precision=0} 
    \tikzsetnextfilename{weak-scaling-comm-2}
    \begin{tikzpicture}[scale=.8]
      \begin{semilogxaxis}[log basis x=2, xmin=1, ymin=0, ymax=40, enlargelimits=.08,
        xmax=128, xtick=data,
        xlabel={Number of Nodes ($P$)},
        xticklabel=\pgfmathparse{2^\tick}\pgfmathprintnumber{\pgfmathresult},
        ylabel={Communication time [\% of execution time]}, major grid style={dotted},ymajorgrids = true,
        legend cell align=left,
        legend style={at={(.5, 1.0)}, anchor=south, draw=none},
        legend columns = 4,
        height=0.45\textwidth,
        width=0.5\textwidth,
        ]

         \addplot[color=black, mark=square] table [x=procs, y expr={\thisrow{commpart}}] {new_timings/Q02_QV1.txt};
        \addlegendentry{Q2};

          \addplot[color=gray, mark=square*] table [x=procs, y expr={\thisrow{commpart}}] {timings/q03-weak-scale-qv2.time};
        \addlegendentry{Q3};
        \addplot[color=gray, mark=*] table [x=procs, y expr={\thisrow{commpart}}] {timings/q03-weak-scale-qv1.time};
        \addlegendentry{Q3 (repl.)};
        \addplot[color=gray, mark=triangle*] table [x=procs, y expr={\thisrow{commpart}}] {new_timings/Q03_QV4.txt};
        \addlegendentry{Q3 (lazy)};

        \addplot[color=black, mark=triangle*] table [x=procs, y expr={\thisrow{commpart}}] {timings/q04.time};
        \addlegendentry{Q4};

		\addplot[color=black, mark=*] table [x=procs, y expr={\thisrow{commpart}}] {new_timings/Q05_QV1.txt};
        \addlegendentry{Q5};

        \addplot[color=black, mark=square*] table [x=procs, y expr={\thisrow{commpart}}] {new_timings/Q11_QV2.txt};
        \addlegendentry{Q11};
      \end{semilogxaxis}
    \end{tikzpicture}
  \caption{Experimental results ($SF=100\cdot P$) - communication}
  \label{fig:eval:communication}
\end{figure}
For the first versions of queries 3 and 21, we evaluate a filter attribute in the first step within a sub-query.
The intermediate result size depends linearly on the scale factor and thus the running time increased.
In a second sub-query, the redistributed intermediate results were joined during the actual aggregation.
In this context, we expected the increasing running time for query 3 and a part of the increased running time for query 21 with growing communication costs because of a doubled intermediate result size for a doubled scale factor.
For the second version of query 3 (lazy) and 21 (late), we execute the query without the remote filter and request the results for required keys at a later stage. We expected a better scaling behavior for these versions because every node only requests a constant number of filter results in weak scaling experiments and only a proportion of the rows are accessed on the remote join path. We did, however, expect slower runtimes for lower number of nodes because we can't perform the whole aggregation step in one run without the results from the remote join path.

We also test alternative implementations of queries 3 (repl.) and 21 (repl.) where we replicate the remote join attribute to eliminate the remote dependency. \mycomment{\frage{m: hier noch sagen, dass dies in praxis zulaessig ist, da nation von supp/cust sehr oft verwendet?}}
Here, the applied strategy for query 3 resulted in constant running time. This is very fast because only at the end we need one collective reduce communication with a fixed-size to collect the final result set.

In contrast to query 3, query 21 scaled worse and did not provide constant
running times with the replicated join attribute. 
This effect can be explained by a second remote dependency, namely the remote group-by key of the aggregation.
Tuples consisting of group-by key and partial aggregate are merged and aggregated by using a custom reduce operator.
The number of partial results increases with the scale factor and, therefore, this operation clearly dominates the running time for larger $P$.
We did not apply our top-$k$ selection by value approximation for query 21 because the integer words of the partial sums are very small.

Overall, we see that the scaling behavior for execution plans that request
filter results explicitly (queries 2, 3 (lazy), 5, 13, 14) is considerably
better than for execution plans that exchange a global filter result over all
nodes (queries 3, 11, 21). This can be accounted to the increasing communication
time required for the allgather operation when adding more data and nodes.
However, for lower number of nodes we observe faster running times when
exchanging a full bitset due to faster local processing.

For queries 5, 13 and 14, a considerable amount of time is spent sorting keys before sending them to other nodes resulting in long overall execution times. The first reason for this is to construct the individual messages, however, this can be avoided by the use of simple indexes. The second reason for sorting is for better compression rates. We decided to accept this loss in runtime because for slower networks than ours, we assume faster execution times when keeping the message sizes as low as possible.
\mycomment{
For query 15, we observed a growing running time with increased weakly scaled factors.
The effect is caused by partial results, which need to be exchanged between the nodes to select the maximum.
In particular, the number of partial results increases linearly with the scale factor.
The plot was created from measures of query 15 with the implementation of our new top-$k$ estimation algorithm for distributed results.
A more precise discussion of query 15 follows.
}
\subsubsection{Top-$k$ Selection}
\label{sec:exp-top-k}
Figure~\ref{fig:eval:no_remote_agg} shows the running times for query 15 using our top-$k$ value approximation algorithm.
Because of weakly-scaled factors, the number of intermediate results doubles in every step and leads to a growing query running time.
We evaluate our algorithm (see Section~\ref{sec:topk}) more precisely by comparing three different implementations of query 15.
We implemented the following variants:
\begin{enumerate}
\item a simple implementation which communicates the full values ($64$ bit required for each) of all partial sums using the library-provided all-to-all algorithm
\item a simple implementation similar to 1) but using the 1-factor algorithm
\item an implementation which uses our top-$k$ solution with approximated values (8 bit approximation).
\end{enumerate}

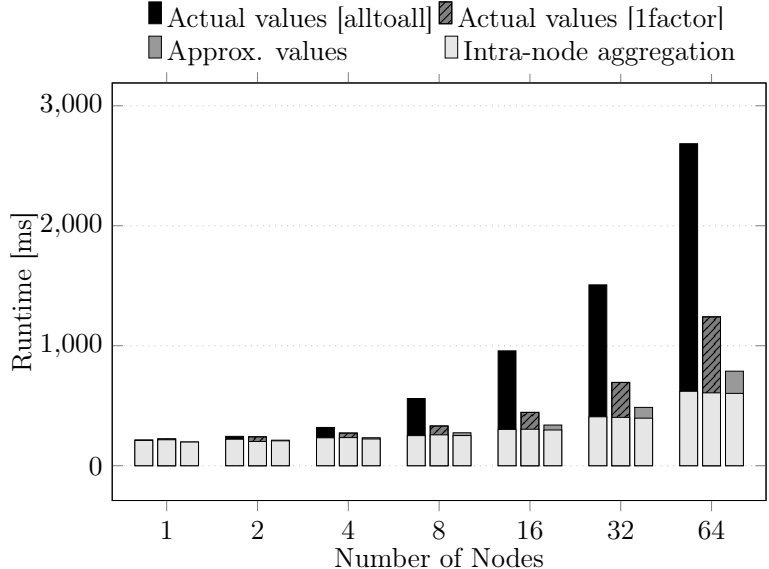
\begin{figure}[!t]
\pgfplotsset{/pgfplots/ybar legend/.style={
    /pgfplots/legend image code/.code={\draw[##1,/tikz/.cd,bar
      width=5pt,yshift=-0.2em,bar shift=0pt] plot coordinates {(0cm,0.8em)
      };},
  }}
  \centering
\tikzexternaldisable
  \begin{tikzpicture}[scale=0.95]
    \begin{axis}[
      bar width=0.25cm,
      height=.45\textwidth,
      width=.65\textwidth,
      major grid style={dotted},
      ybar, ymax=2900, ymin=0, ymajorgrids = true,
      xtick=data, xticklabels from table={timings/query_benchmark_15-2.time}{Label},
      legend cell align=left, legend style={at={(.5,1.02)}, anchor=south, name={leg1}, draw=none},
      enlargelimits=.1,
      tick label style=white, tick style={color=white},
      legend columns = 2
      ]
      \addplot [fill=black] table [ y = Runtime, x expr=\coordindex, meta=Variant] {timings/query_benchmark_15-1.time};
      \addlegendentry{Actual values [alltoall]}
      \addplot [fill=black!50!white, postaction={pattern=north east lines}] table [ y = Runtime, x expr=\coordindex, meta=Variant] {timings/query_benchmark_qv1_1factor.time};
      \addlegendentry{Actual values [1factor]}
      \addplot [restrict x to domain={0:6}, fill=black!40!white] table [ y = Runtime, x expr=\coordindex, meta=Variant] {timings/query_benchmark_15-2.time};
      \addlegendentry{Approx. values}
    \end{axis}
    \begin{axis}[ bar width=0.25cm,
      height=.45\textwidth,
      width=.65\textwidth,
      ybar, ymax=2900, ymin=0, ylabel={Runtime [ms]},
      xtick=data, xlabel={Number of Nodes}, xticklabel pos=left, xticklabels from table={timings/query_benchmark_15-2.time}{Label},
      legend cell align=right, legend style={at={(.735,1.02)}, anchor=south , name={leg2}, draw=none},
      enlargelimits=.1
      ]
      \addplot [fill=black!10!white] table [y expr={\thisrow{Runtime}-\thisrow{Commtime}}, x expr=\coordindex, meta=Variant] {timings/query_benchmark_15-1.time};
      \addlegendentry{Intra-node aggregation}
      \addplot [fill=black!10!white] table [y expr={\thisrow{Runtime}-\thisrow{Commtime}}, x expr=\coordindex, meta=Variant] {timings/query_benchmark_qv1_1factor.time};
      \addplot [restrict x to domain={0:6}, fill=black!10!white] table [y expr={\thisrow{Runtime}-\thisrow{Commtime}}, x expr=\coordindex, meta=Variant] {timings/query_benchmark_15-2.time};
    \end{axis}
  \end{tikzpicture}
  \caption{Q15 -- Actual and approximated values ($SF=100\cdot P$)}
  \label{fig:eval:approx_vs_trivial}
\end{figure}

The results of our experiments with a weakly scaled factor ($SF=100\cdot P$) can be seen in Fig.~\ref{fig:eval:approx_vs_trivial}, where all three bars are clustered and relate to the same number of nodes.
A bar represents either the simple solution (every first two bars, black with MPI all-to-all and dark gray with 1-factor) or our implemented top-$k$ algorithm with value approximation (every third bar, gray).
Light gray parts identify the time used for the local aggregation and they are expected to be equal among the three experiments.

\mycomment{
We described an improvement for the personalized all-to-all communicator with the 1-factor algorithm in Section~\ref{sec:tuning-basic-communication-functions}.
In this experiment, we measured its performance against the library implementation (Open MPI v1.6.3) for the simple variant (communicate all partial sums).
As we observe from the experimental results, the 1-factor always has a lower running time for $P>2$ and outperforms the library implementation of the personalized all-to-all.
\frage{j@m: can you rewrite this?: Our first observation is a strong variation related to the MPI all-to-all throughput (of Open MPI v1.6.3 library), which ranges between $500$\unitfrac{MB}{s} and $1.5$\unitfrac{GB}{s}.}\antwort{m: done. the strong-variation fact not mentioned because in Section III.}
}
First, the 1-factor implementation requires less communication time for the same amount of data as the library-provided all-to-all algorithm for $P>2$.
Second, we compare the simple variants with the top-$k$ algorithm.\frage{j was: Our second observation is related to the implemented algorithms.}
We predicted lower running times for the approximative algorithm \emph{(gray bar)} due to a factor $8$ less data to be exchanged -- compared to exchanging the actual values ($64\unit{bit}$ keys originally, $8\unit{bit}$ for encoded values).
For better comparison, we also used the 1-factor algorithm to exchange the encoded values.
The overhead of encoding and decoding the partial sums requires computation time as well, but we parallelized it using multi-threading.
Moreover, the intra-node throughput with $14\unitfrac{GB}{s}$\mycomment{\footnote{$0.072\unit{s}$ walltime and $1.890\unit{s}$ user time $\rightarrow 2819.4\%$ CPU usage on 16 explicit cores per node.}} for encoding and $4\unitfrac{GB}{s}$ for decoding (the decoding includes the required aggregation of partial sums per key) are higher than the specified point-to-point network throughput of $3700\unitfrac{MB}{s}$.
Our prediction for the top-$k$ algorithm with partial results approximation was correct by observing speedups up to $2.3$ over the simple approach (with 1-factor).

\subsection{Intra-node parallelism}

A further experiment allows evaluating the effect of intra-node parallelism on query running times.
Note that each cluster node contains $16$ physical cores and Hyper-Threading is enabled.
We used weakly-scaled factors and run the queries on $128$ nodes.
Next, the relative speedups of the weakly-scaled experiments with enabled multi-threading over the single-threaded running times were calculated for each query.
The speedups are shown in Table~\ref{tbl:singlecore}.
The queries which require little communication achieve high speedup of 18--24, even more than the factor 16 to be expected from the number of cores. The speedup is lower for the communication-bound queries. But even there we achieve speedups of up to 6.

\begin{table}
  \centering
  \renewcommand{\arraystretch}{1.2} 
  \caption{Speedup of intra-node parallelism ($128$ nodes)}
  \label{tbl:singlecore}
  \begin{tabular}{| l |  r || l | r |}
    \hline
    Query & \multicolumn{1}{c ||}{Speedup} & Query & \multicolumn{1}{c |}{Speedup}  \\ \hline
    1 & 18.7 & 13 & 4.7  \\
    2 & 2.5 & 14 & 6.0 \\
    3 & 5.9  & 15 & 3.1 \\
    3 (lazy) & 8.2 & 18 & 24.2  \\
    4 & 18.1 & 21 & 5.7  \\
    5 & 6.6 & 21 (late) & 5.9 \\
    11 & 1.8 & & \\
    \hline
  \end{tabular}
\end{table}
\subsection{Comparison with TPC-H Record Holder}

\begin{table}[t]
\centering
\caption{Power test, our system and current record holder. SPEC values are SPEC$_{\text{intrate}}$ from \cite{spec:2006}.}
\label{tbl:records}}{
\centering
\begin{tabular}{|l|r|r|r|r|r|r|r|r|}
    \hline
    & \multicolumn{5}{c|}{$SF=10\,000$} & \multicolumn{3}{c|}{$SF=30\,000$} \\
    \hline
    \multirow{2}{*}{Query} & \multicolumn{1}{c|}{We} &\multicolumn{2}{c|}{EXASol 4.0} & \multicolumn{2}{c|}{EXASol 5.0} & \multicolumn{1}{c|}{We} & \multicolumn{2}{c|}{EXASol 5.0} \\ 
    & \multicolumn{1}{c |}{in $[\unit{s}]$} & \multicolumn{1}{c }{in $[\unit{s}]$} & \multicolumn{1}{c|}{factor} & \multicolumn{1}{c }{in $[\unit{s}]$} & \multicolumn{1}{c|}{factor}& \multicolumn{1}{c|}{in $[\unit{s}]$}& \multicolumn{1}{c }{in $[\unit{s}]$} & \multicolumn{1}{c|}{factor} \\
    \hline
    1 & 0.442 & 10.6 & 24.0 & 8.1 & 18.3 &0.625 & 20.7 & 33.1 \\
    2 & 0.063 & 1.1 & 17.5 & 0.9 & 14.3 &0.093 & 2.0 & 21.5\\
    3  & 0.945 & 6.9 & 7.3 & 6.7 & 7.1 &2.786  & 16.0 & 5.7\\
    3lazy & 0.610 & 6.9 & 11.3 & 6.7 & 11.0&0.867  & 16.0 & 18.5\\
    4 & 0.137 & 1.8 & 13.1 & 1.8 & 13.1 &0.124 & 4.1 & 33.0\\
    5 & 2.539 & 7.2 & 2.8 & 4.2 &  1.7&1.463 & 11.5 & 7.9 \\
    11 & 0.404 & 15.0 & 37.1 & 12.1 & 30.0&0.688 & 35.6 & 51.7\\
    13 & 6.833 & 8.8 & 1.3 & 7.8 & 1.1&4.548 & 21.1 & 4.6\\
    14 & 1.091 & 2.7 & 2.5 & 3.0 & 2.7&1.659 & 7.9 & 4.8\\
    15 & 1.156 & 10.8 & 9.3 & 11.4 & 9.9 &3.331 & 29.9 & 9.0 \\
    18 & 0.212 & 10.8 & 50.9 & 11.9 &  56.1&0.301 & 30.5 & 101.3 \\
    21 & 1.122 & 30.6 & 27.2 & 3.9 &  3.5&2.306 & 10.3 & 4.5\\ 
    21late & 0.869 & 30.6 & 35.2 & 3.9 & 4.5&1.501 & 10.3 & 6.9\\
   \hline\hline
   SPEC& 625 & 419 & 0.7 & 827 & 1.3&625 & 827 & 1.3\\ 

   Nodes& 60 & 60 & 1.0 & 34 & 0.6& 128 & 40 & 0.3\\
   RAM& $3.8\unit{TB}$ & $4.3\unit{TB}$ & $1.1$ & $5.4\unit{TB}$ & 1.4 & $8.2\unit{TB}$ & $12.8\unit{TB}$ & $1.6$\\ 
   \hline
  \end{tabular}
\end{table}  
We execute an additional test series with $SF=10\,000$ on $60$ nodes and with $SF=30\,000$ on $128$ nodes in order to compare our results to the current \mbox{TPC-H} record holder.
The current record holder for Scalefactors $10\,000,
30\,000$ and $100\,000$ is EXASolution 5.0 on a Dell PowerEdge 720xd. Each
machine has two Intel Xeon E5-2680v2 10C $2.8\unit{GHz}$ processors with 10
cores per chip. \footnote{State of August 29, 2017}

In the official results for $SF=10\,000$, EXASolution 5.0 is run on 34 Dell PowerEdge R720xd nodes. However, we cannot run our implementation on only 34 with $SF=10\,000$ because our system does not have enough memory.
Thus, we also provide a comparison to the second best result for $SF=10\,000$ which is EXASolution 4.0 an a Dell PowerEdge 710 with 60 nodes.
Each node has $72\unit{GB}$ RAM and they use two Intel Xeon X5690 QC $3.46\unit{GHz}$, each chip with six cores.
Both systems contain 60 nodes.
The total RAM of the EXASol cluster is $4320\unit{GB}$ whereas our cluster has $3840\unit{GB}$ of RAM available.
The interconnection between the nodes is realized by an InfiniBand $4$X QDR network, which is the same as in our cluster.

The results are provided in Table~\ref{tbl:records}, where we show for each query: our running time, the running time of EXASol, and the factor by which we are faster than the competitor.
As can be seen, the running times of our implementation are better by a factor of up to $50$ compared to EXASolution 4.0. The comparison with EXASolution 5.0 is only of limited significance because of the use of less machines. At the one hand less nodes naturally yield less parallelism, one the other hand the communication overhead is reduced because more work can be done locally. Compared to EXASolution 5.0 our results are better by a factor of up to $56$ for $SF=10\,000$ and up to 101 for $SF=30\,000$.

As our cluster uses different machines than EXASol we provided SPEC$_{\text{intrate}}$ numbers of the SPEC 2006 benchmark \cite{henning2006spec} for comparison \footnote{The TPC-H benchmark for EXASolution 5.0 was run on a Dell PowerEdge 720xd, whereas the SPEC$_{\text{intrate}}$ benchmark was run on a Dell PowerEdge 720. The two models differ in the maximum number of internally mounted disks, which is not of interest for our comparison. Also, the memory configuration was different for the two benchmarks.}.

\section{Conclusion and Future Work}\label{sec:conclusion}


We have demonstrated that distributed query execution using message
passing in the combination with intra-node shared-memory parallelism
can be performed very efficiently in a cluster.    \mycomment{We refined the
  distributed query classification of Akal et al.\ \cite{akal2002olap}
  with remote data classes.  In contrast to their classification we
  can process queries with remote data dependency without data
  replication. }\frage{mw: removed classific.}
We developed several techniques for resolving remote data dependency by using efficient communication algorithms.
Moreover, we demonstrated the application of our concepts on a subset of TPC-H benchmark queries.
The evaluation showed that we are able to query large amounts of data with short response times using a cluster and combining sophisticated collective operations (from MPI), multi-threading and efficient algorithms.
In particular, we efficiently implemented clustered SQL query execution with
data sets of up to $30\,000\unit{GB}$ of uncompressed data in main memory and achieved query running times with a factor of one to two orders of magnitudes faster than one of the best results reported for clustered execution of TPC-H on scale factor $10\,000$. 

\paragraph*{Future Work}
In this work, the individual query plans with its physical operators are chosen manually and then translated into static C++ functions.   To allow the execution of arbitrary SQL queries the system must be enhanced by two components:   First, a cost based optimizer having all information at hand to choose a cost optimal query plan.   Second, a compiler translating the query plan into native code parts, such that the resulting executable code   is equal or similar to the C++ functions in this work.   With this, the additional execution time to allow dynamic arbitrary queries is mainly the time spent in these two components.   We are currently working on a productive database engine which already features a cost based optimizer and a query compiler   compiling incoming SQL queries into native code in the range of centiseconds. Using data volumes and queries as in our evaluation, running times are dominated by the execution and not the compilation of plans.   Although not all algorithms and plans described in this work are yet integrated into the productive system, we do not see any obstacle for doing so.   Also, there are other systems demonstrating the feasibility of such an approach with similar observations (see also Section~\ref{sec:related})

For larger systems with thousands of nodes, fault tolerance will become important because node failures and other errors will be common place.
The challenge here is to introduce some redundancy without excessive cost.
\mycomment{\review{ps: drop the GPU stuff? this is currently not interesting in particular for big data. j@ps: you are right, i dropped it. actually it was motivated by adding a reference.}A third field of research is the usage of heterogeneous clusters with CPU and GPU for a parallel query processing, addressed by Rauhe et al.\ \cite{rauheadbis2013}.
A major problem with GPUs is the limited bandwidth of the communication bus between main memory and the GPUs memory.
Their proposed solution can also be applied in order to take advantage of potentially available GPU chipsets on each node and to gain further speedups for distributed query processing.}

We applied range partitioning and co-partitioning for specific tables because TPC-H uses synthetic data and we could achieve a good load balance with those strategies.
Nevertheless, there are also adaptive partitioning methods like \cite{stohr2000multi}, adjusting the partitioning according to the current access patterns and workload.
Finally, we evaluated our implementation by using a high performance cluster with a fast InfiniBand network for node interconnection.
An application on commodity hardware with slower networks is also an interesting use case as part of future work.
\mycomment{
\begin{itemize}
\item discuss heterogeneous clusters with CPU + GPU, reference ADBIS 2013 paper
\item automatic plan generation based on our classes
\item fault tolerance - ausfallsicherheit wenn eine node ausfaellt (redundants), wird relevant ab $\gg$1000 maschinen
\item integrate mixed workload OLTP/OLAP into cluster / transactions, but it should be orthogonal (referenzen zu andren sachen)
\item application for commodity hardware/network (via internet?), slow networks
\end{itemize}
}



\section*{Acknowledgments}
The authors would like to thank Franz F\"arber, David
Kernet, Norman May, Ingo M\"uller and Sebastian Schlag for helpful
suggestions and comments.  \frage{j, reordered by alphabet (may before
 mueller)}

\bibliographystyle{plain}
\bibliography{references}

\begin{thebibliography}{10}

\bibitem{abadi2008column}
Daniel~J Abadi, Samuel~R Madden, and Nabil Hachem.
\newblock Column-stores vs. row-stores: How different are they really?
\newblock In {\em Proc. of SIGMOD}, pages 967--980. ACM, 2008.

\bibitem{akal2002olap}
Fuat Akal, Klemens B{\"o}hm, and Hans-J{\"o}rg Schek.
\newblock {OLAP} query evaluation in a database cluster: a performance study on
  intra-query parallelism.
\newblock In {\em ADBIS}, pages 218--231. Springer, 2002.

\bibitem{bader2002algorithm}
David~A Bader, Bernard~ME Moret, and Peter Sanders.
\newblock Algorithm engineering for parallel computation.
\newblock In {\em Experimental Algorithmics}, pages 1--23. Springer, 2002.

\bibitem{bruck1997efficient}
Jehoshua Bruck, Ching-Tien Ho, Shlomo Kipnis, Eli Upfal, and Derrick
  Weathersby.
\newblock Efficient algorithms for all-to-all communications in multiport
  message-passing systems.
\newblock {\em Parallel and Distributed Systems, IEEE Transactions on},
  8(11):1143--1156, 1997.

\bibitem{poly3}
I.~J. Bush and W.~Smith.
\newblock {The Weak Scaling of DL\_POLY 3}, 2013.
\newblock Accessed on May 28, 2013.

\bibitem{cao2004efficient}
Pei Cao and Zhe Wang.
\newblock Efficient top-k query calculation in distributed networks.
\newblock In {\em Proc. of PODC}, pages 206--215. ACM, 2004.

\bibitem{chockler2001group}
Gregory~V Chockler, Idit Keidar, and Roman Vitenberg.
\newblock Group communication specifications: a comprehensive study.
\newblock {\em ACM Comput. Surv.}, 33(4):427--469, 2001.

\bibitem{cuzzocrea2013olap}
Alfredo Cuzzocrea, Rim Moussa, and Guandong Xu.
\newblock Olap*: Effectively and efficiently supporting parallel olap over big
  data.
\newblock In {\em Model and Data Engineering}, pages 38--49. Springer, 2013.

\bibitem{dees2013}
Jonathan Dees and Peter Sanders.
\newblock Efficient many-core query execution in main memory column-stores.
\newblock {\em ICDE}, pages 350--361, 2013.

\bibitem{dewitt1992}
David DeWitt and Jim Gray.
\newblock Parallel database systems: The future of high performance database
  systems.
\newblock {\em Commun. ACM}, 35(6):85--98, 1992.

\bibitem{dewitt1990parallel}
David~J DeWitt and Jim Gray.
\newblock Parallel database systems: The future of database processing or a
  passing fad?
\newblock {\em ACM SIGMOD Record}, 19(4):104--112, 1990.

\bibitem{eavis2010parallel}
Todd Eavis, George Dimitrov, Ivan Dimitrov, David Cueva, Alex Lopez, and Ahmad
  Taleb.
\newblock Parallel {OLAP} with the sidera server.
\newblock {\em Future Gener. Comput. Syst.}, 26(2):259--266, 2010.

\bibitem{fagin2001optimal}
Ronald Fagin, Amnon Lotem, and Moni Naor.
\newblock Optimal aggregation algorithms for middleware.
\newblock In {\em Proc. of PODS}, pages 102--113. ACM, 2001.

\bibitem{freedman2014compilation}
Craig Freedman, Erik Ismert, and Per-{\AA}ke Larson.
\newblock Compilation in the microsoft sql server hekaton engine.
\newblock {\em IEEE Data Eng. Bull.}, 37(1):22--30, 2014.

\bibitem{furtado2005physical}
Camille Furtado, Alexandre~AB Lima, Esther Pacitti, Patrick Valduriez, and
  Marta Mattoso.
\newblock Physical and virtual partitioning in olap database clusters.
\newblock In {\em Computer Architecture and High Performance Computing, 2005.
  SBAC-PAD 2005. 17th International Symposium on}, pages 143--150. IEEE, 2005.

\bibitem{gao2005consistent}
Bin Gao, Tie-Yan Liu, Xin Zheng, Qian-Sheng Cheng, and Wei-Ying Ma.
\newblock Consistent bipartite graph co-partitioning for star-structured
  high-order heterogeneous data co-clustering.
\newblock In {\em Proc. of SIGKDD}, pages 41--50. ACM, 2005.

\bibitem{gabriel2004open}
Richard~L Graham, Timothy~S Woodall, and Jeffrey~M Squyres.
\newblock Open {MPI}: A flexible high performance {MPI}.
\newblock In {\em Parallel Processing and Applied Mathematics}, pages 228--239.
  Springer, 2006.

\bibitem{gropp1999using}
William Gropp, Ewing~L Lusk, and Anthony Skjellum.
\newblock {\em Using MPI: Portable Parallel Programming with the
  Message-Passing Interface}.
\newblock MIT press, 1999.

\bibitem{gustafson1988reevaluating}
John~L Gustafson.
\newblock Reevaluating {A}mdahl's law.
\newblock {\em Commun. ACM}, 31(5):532--533, 1988.

\bibitem{han2007progressive}
Wook-Shin Han, Jack Ng, Volker Markl, Holger Kache, and Mokhtar Kandil.
\newblock Progressive optimization in a shared-nothing parallel database.
\newblock In {\em Proceedings of the 2007 ACM SIGMOD international conference
  on Management of data}, pages 809--820. ACM, 2007.

\bibitem{henning2006spec}
John~L Henning.
\newblock {SPEC CPU2006} benchmark descriptions.
\newblock {\em ACM SIGARCH Computer Architecture News}, 34(4):1--17, 2006.

\bibitem{HubSan16}
Lorenz H{\"{u}}bschle{-}Schneider and Peter Sanders.
\newblock Communication efficient algorithms for top-$k$ selection problems.
\newblock In {\em 30th IEEE International Parallel {\&} Distributed Processing
  Symposium (IPDPS)}, 2016.

\bibitem{karanasos2014dynamically}
Konstantinos Karanasos, Andrey Balmin, Marcel Kutsch, Fatma Ozcan, Vuk
  Ercegovac, Chunyang Xia, and Jesse Jackson.
\newblock Dynamically optimizing queries over large scale data platforms.
\newblock In {\em Proceedings of the 2014 ACM SIGMOD international conference
  on Management of data}, pages 943--954. ACM, 2014.

\bibitem{kemper2011hyper}
A.~Kemper and T.~Neumann.
\newblock {{HyPer}: A hybrid {OLTP}\&{OLAP} main memory database system based
  on virtual memory snapshots}.
\newblock In {\em Proc. of ICDE}, pages 195--206. IEEE, 2011.

\bibitem{krikellas2010generating}
Konstantinos Krikellas, Stratis~D Viglas, and Marcelo Cintra.
\newblock Generating code for holistic query evaluation.
\newblock In {\em Proc. of ICDE}, pages 613--624. IEEE, 2010.

\bibitem{Lee2013HANA}
Juchang Lee, Yong~Sik Kwon, F.~F{\"a}rber, M.~Muehle, Chulwon Lee, C.~Bensberg,
  Joo~Yeon Lee, A.H. Lee, and W.~Lehner.
\newblock Sap hana distributed in-memory database system: Transaction, session,
  and metadata management.
\newblock In {\em Data Engineering (ICDE), 2013 IEEE 29th International
  Conference on}, pages 1165--1173, April 2013.

\bibitem{LemBoy15}
Daniel Lemire and Leonid Boytsov.
\newblock Decoding billions of integers per second through vectorization.
\newblock {\em Software Practice \& Experience}, 45(1), 2015.

\bibitem{lima2004adaptive}
Alexandre~AB Lima, Marta Mattoso, and Patrick Valduriez.
\newblock Adaptive virtual partitioning for {OLAP} query processing in a
  database cluster.
\newblock In {\em Proc. SBBD}, pages 92--105, 2004.

\bibitem{lima2004olap}
Alexandre~AB Lima, Marta Mattoso, and Patrick Valduriez.
\newblock {OLAP} query processing in a database cluster.
\newblock In {\em Euro-Par 2004 Parallel Processing}, pages 355--362. Springer,
  2004.

\bibitem{nagel2014code}
Fabian Nagel, Gavin Bierman, and Stratis~D Viglas.
\newblock Code generation for efficient query processing in managed runtimes.
\newblock {\em Proc. of the VLDB}, 7(12):1095--1106, 2014.

\bibitem{neumann2011efficiently}
Thomas Neumann.
\newblock Efficiently compiling efficient query plans for modern hardware.
\newblock {\em Proc. of VLDB}, 4(9):539--550, 2011.

\bibitem{pavlo09}
Andrew Pavlo, Erik Paulson, Alexander Rasin, Daniel~J. Abadi, David~J. DeWitt,
  Samuel Madden, and Michael Stonebraker.
\newblock A comparison of approaches to large-scale data analysis.
\newblock In {\em Proc. of SIGMOD}, pages 165--178. ACM, 2009.

\bibitem{Poess:tpch}
Meikel Poess and Chris Floyd.
\newblock New {TPC} benchmarks for decision support and web commerce.
\newblock {\em SIGMOD Rec.}, 29(4):64--71, 2000.

\bibitem{reinders2010intel}
James Reinders.
\newblock {\em Intel threading building blocks: outfitting C++ for multi-core
  processor parallelism}.
\newblock O'Reilly Media, 2010.

\bibitem{rodiger2015high}
Wolf R{\"o}diger, Tobias M{\"u}hlbauer, Alfons Kemper, and Thomas Neumann.
\newblock High-speed query processing over high-speed networks.
\newblock {\em Proc. of the VLDB}, 9(4):228--239, 2015.

\bibitem{sanders2002:1factor}
Peter Sanders and Jesper~Larsson Tr{\"a}ff.
\newblock The hierarchical factor algorithm for all-to-all communication.
\newblock In {\em Euro-Par 2002 Parallel Processing}, pages 799--803. Springer,
  2002.

\bibitem{Shute2013F1}
Jeff Shute, Radek Vingralek, Bart Samwel, Ben Handy, Chad Whipkey, Eric
  Rollins, Mircea Oancea, Kyle Littlefield, David Menestrina, Stephan Ellner,
  John Cieslewicz, Ian Rae, Traian Stancescu, and Himani Apte.
\newblock F1: A distributed sql database that scales.
\newblock {\em Proc. VLDB Endow.}, 6(11):1068--1079, August 2013.

\bibitem{spec:2006}
{SPEC}.
\newblock {All SPEC CPU2006 Results Published by SPEC}, 2013.
\newblock Accessed on May 17, 2013.

\bibitem{stohr2000multi}
Thomas St{\"o}hr, Holger M{\"a}rtens, and Erhard Rahm.
\newblock Multi-dimensional database allocation for parallel data warehouses.
\newblock In {\em Proc. of VLDB}, pages 273--284, 2000.

\bibitem{council2008tpc}
TPC.
\newblock {TPC Benchmark H v2.14.4}, 2012.

\bibitem{valduriez1993parallel}
Patrick Valduriez.
\newblock Parallel database systems: open problems and new issues.
\newblock {\em Distributed and parallel Databases}, 1(2):137--165, 1993.

\bibitem{WDS13}
Martin Weidner, Jonathan Dees, and Peter Sanders.
\newblock Fast {OLAP} query execution in main memory on large data in a
  cluster.
\newblock In {\em IEEE Int. Conf. on Big Data}, pages 518--524, 2013.

\end{thebibliography}
%



\begin{appendix}
\section{Unimplemented TPC-H Queries}\label{app:notImplemented}
\begin{description}
\item[6] Trivial local query.
\item[7] Two remote filters (on \Id{customer} and \Id{supplier} nations).  Planning is interesting: First query one of the remote filters.  The result size determines which strategy is best for the second filter.
Aggregation is cheap since the result is just 4 lines (2 years $\times$ 2 nations pairs).
\item[8] Three remote filters. Similar issues as for Query 7.
\item[9] Very expensive query. We have a remote data dependency on
  \Id{ps\_supplycost} and \Id{supplier} nation. The only filter condition is on
  colors which could be supported by a full text index.
  The filter condition can be evaluated remotely and we only send the
  resulting supply costs. Aggregation is cheap once more.
\item[10] Rather selective local filters. We are aggregating on a   remote key and thus might profit from the top-k selection strategies
  described in Section~\ref{sec:topk}.
\item[12] Trivial local query.
\item[16] Very selective local filter. Hence variant one of Section~\ref{sec:filter_remote} is promising.
\item[17] Filtering by remote attribute. No local filter.
\item[19] Highly selective local filter. Hence Variant 1 from Section~\ref{sec:filter_remote}  is always best and we can expect good scalability.
  There are remote filters on \Id{part}. Aggregation of local data to a
  single value.
\item[20] Rather complex with aggregation on a nonlocal key as
  a subquery. Several remote filter conditions.
\item[22] Can be executed almost locally if orders and customers are
  copartitioned or if we have an index mapping customers to orders.
\end{description}
\end{appendix}

\end{document}